# Designing layered 2D skyrmion lattices in moiré magnetic heterostructures

Bilal Jabakhanji[1] and Doried Ghader*[1]

[1] College of Engineering and Technology, American University of the Middle East, Egaila 54200, Kuwait

## 1. Abstract

Skyrmions are promising for the next generation of spintronic and magnonic devices, but their zero-field stability and controlled nucleation through chiral interactions remain challenging. In this theoretical study, we explore the potential of moiré magnetic heterostructures to generate ordered skyrmion lattices from the stacking-dependent magnetism in 2D magnets. We consider heterostructures formed by twisting ultrathin $CrBr_3$ films on top of $CrI_3$ substrates, assuming a moderate interfacial Dzyaloshinskii-Moriya interaction. At large moiré periodicity and appropriate substrate thickness, a moiré skyrmion lattice emerges in the interfacial $CrBr_3$ layer due to the weaker exchange interactions in $CrBr_3$ compared to $CrI_3$. This lattice is then projected to the remaining layers of the $CrBr_3$ film via emergent chiral interlayer fields. By varying the pristine stacking configurations within the ultrathin $CrBr_3$ film, we realize layered ferromagnetic and antiferromagnetic skyrmion lattices without the need for a permanent magnetic field. Our findings suggest the possibility of creating colorful skyrmion lattices in moiré magnetic heterostructures, enabling further exploration of their fundamental properties and technological relevance.

## 2. Introduction

Topology continues to reform condensed matter physics and material science, offering vast opportunities for fundamental discoveries and technological applications. In magnetism, topological spin textures (TSTs) such as skyrmions are the subject of intense research as they hold substantial potential for future topological spintronic and magnonic devices[1–6]. The most common source of TSTs is the chiral nearest neighbor (NN) Dzyaloshinskii-Moriya interaction (DMI), which arises in non-centrosymmetric magnets without space-inversion symmetry[7–14]. Recently, research on TSTs extended to the newly discovered 2D magnets [15,16], opening new horizons for topological magnetism in the 2D limit[17–25]. The quest for 2D TSTs has motivated a search for novel mechanisms for their stabilization. In particular, moiré skyrmionics emerged as a new research field aiming to create TSTs through the manipulation of competing interactions in moiré superlattices[26]. In a 2D magnetic bilayer, a moiré superlattice emerges from lattice mismatches or twisting, generating a spatially modulated interlayer coupling within a moiré supercell. At large moiré periodicity, the interlayer interaction can dominate over the intralayer coupling, creating exotic magnetic phases with non-collinear spin textures[26–33].

Theoretical studies[26,28,34–36] on moiré magnets predicted the emergence of TSTs, assisted by the moiré interlayer fields. When the NN DMI is negligible, the vorticity and helicity of the TSTs act as degrees of freedom, resulting in multi-flavored TSTs[30,34]. A sizeable NN DMI, however, locks the degrees of freedom, producing Néel-type moiré skyrmions[30,31]. Recent experiments[29,32,33,37] have reported non-collinear magnetic states in moiré $CrI_3$ superlattices. However, limitations in the experimental techniques have not yet allowed for a full exploration of the possible topological nature of moiré spin textures.

The 2D chromium trihalide magnets $CrX_3$ ($X = I, Br$) have received exceptional attention due to their intriguing stacking-dependent magnetism and topological spin excitations. Theoretical and experimental investigations on $CrI_3$ and $CrBr_3$ reported topological band gaps in their magnonic spectra [38–42], indicating the presence of strong NN Kitaev or next NN DMI interactions in these materials. Meanwhile, the interlayer exchange interaction in these materials can exhibit either ferromagnetic (FM) or antiferromagnetic (AFM) behavior, depending on the stacking configuration. More precisely, the interlayer interaction in $CrI_3$ ($CrBr_3$) is AFM in the monoclinic (R-type) stacking and FM in the rhombohedral (H-type) stacking[31,34,43–46], respectively. The stacking-dependent magnetism motivated intensive investigations of moiré TSTs in twisted $CrI_3$ and $CrBr_3$ homo-bilayers[29–32,34,47]. More recently, a theoretical work predicted stacking-dependent magnetism in $CrBr_3/CrI_3$ hetero-bilayers[48], but their moiré ground states are not yet studied.

Magnetic hetero-bilayers display rich stacking-dependent magnetism[26,48–50], but their moiré magnetic phases remain widely unexplored. So far, theoretical studies on moiré magnetic heterostructures are limited to a specific setup, namely an FM monolayer on an AFM substrate (or vice versa), investigated within the single-layer approximation (the substrate spins are assumed fixed)[26,28,35,51]. Meanwhile, enormous opportunities exist to construct moiré heterostructures from the rapidly expanding family of 2D magnets, generating a vast unexplored arena for moiré-engineered magnetism. Recent experiments on pristine (untwisted) magnetic heterostructures reported exciting magnetic observations like the emergence of the magnetic proximity effect and interfacial NN DMI [52,53]. In particular, Cheng et al. explored the magnetic proximity effect in pristine $CrI_3/CrCl_3$ heterostructures, where the interfacial FM coupling significantly affected the magnetic anisotropy in the $CrCl_3$ layer proximal to $CrI_3$. In another study[53], Wu et al. explored pristine ultrathin $Cr_2Ge_2Te_6/Fe_3GeTe_2$ heterostructures. The broken inversion symmetry at the vdW interface induced sizeable chiral DMI, stabilizing magnetic skyrmions at the two sides of the interface. Nevertheless, no relevant indications of moiré magnetism were observed in the pristine stacking of the studied $CrI_3/CrCl_3$ and $Cr_2Ge_2Te_6/Fe_3GeTe_2$ heterostructures.

Achieving control over skyrmion nucleation and spatial distribution is desired for their potential applications in magnonics and spintronics[54,55]. However, this continues to be a significant challenge in magnetic thin films and multilayers, as DMI-induced skyrmions tend to nucleate randomly and require a permanent magnetic field [54–56]. By contrast, moiré-engineered skyrmions can persist without a magnetic field[30], with well-defined shapes and precise locations within the moiré supercell. The twist angle serves as a knob to control their position, size, and spatial arrangement[25,26,30,31,34]. Therefore, compared to spontaneously nucleated DMI-skyrmions, moiré skyrmions appear to have the potential to achieve spatially ordered skyrmion arrangements that do not require a permanent field.

Nevertheless, the emergence of moiré skyrmion lattices is not straightforward and requires investigation of adequate materials and setups. In magnetic moiré homo-bilayers, skyrmion lattices are unlikely due to the identical magnetic interactions in the top and bottom layers. When accounting for the full spin dynamics, TSTs can form in any layer[30], disrupting the magnetic moiré periodicity in realistic samples. Moiré magnetic heterostructures, which combine materials with different magnetic properties, are candidates for confining skyrmions to specific layers. However, the investigation of their advantages requires modeling beyond the single-layer approximation. The single-layer approximation for moiré bilayers neglects the spin dynamics and intralayer interactions in one of the layers, making it a vague approximation. Moreover, this approximation cannot model moiré architectures with three or more magnetic layers. As the number of layers increases in a moiré setup, confining skyrmions to a specific layer becomes increasingly challenging, since the coupled spin dynamics in all layers cooperate to shape the magnetic ground state from the competing intra and interlayer interactions. These complexities call for improved modeling and simulations capable of demonstrating controlled layer-by-layer skyrmion formation, which is a prerequisite for realizing (layered) 2D skyrmion lattices in moiré magnetic heterostructures.

This work proposes magnetic moiré heterostructures as platforms for designing layered 2D skyrmion lattices, using twisted $(CrBr_3)_n/(CrI_3)_{n'}$ heterostructures as prototypes to demonstrate our findings. The $(CrBr_3)_n/(CrI_3)_{n'}$ heterostructure derived from an ultrathin pristine $CrBr_3$ film ($n$ layers) twisted by an angle $\theta$ relative to an ultrathin pristine $CrI_3$ substrate ($n'$ layers). The heterostructures are further diversified by considering an FM or AFM stacking between adjacent layers in the pristine films. The magnetic ground state is determined using stochastic Landau-Lifshitz-Gilbert simulations (sLLG). These simulations are grounded in theoretical modeling that incorporates the full spin dynamics[30] and accounts for the diverse competing interlayer and intralayer interactions in the hetero-bilayers. Considering the difference in exchange interactions between $CrBr_3$ and $CrI_3$, we adjusted the thicknesses of the films and explored the potential to restrict the formation of skyrmions to specific layers of the heterostructure.

As an initial step, we examine the skyrmion nucleation in moiré $CrBr_3/CrI_3$ hetero-bilayers and observe moiré skyrmion lattices forming explicitly in the $CrBr_3$ layer due to its weaker intralayer exchange compared to $CrI_3$. We determine the threshold interfacial NN DMI (treated as a parameter) that can stabilize the $CrBr_3$ skyrmion lattice for commensurate twist angles between $1.02°$ and $3.15°$.

Subsequently, we explored the formation of $CrBr_3$ layered skyrmion lattices in $(CrBr_3)_2/(CrI_3)_2$ and $(CrBr_3)_3/(CrI_3)_2$ heterostructures. In both heterostructures, the pristine substrate is stacked in an FM configuration, while adjacent $CrBr_3$ layers are stacked in a pristine FM or AFM configuration. In these setups, the moiré interlayer interaction produces a skyrmion lattice in the interfacial $CrBr_3$ layer. The spin textures in the interfacial layer cooperate with the uniform $CrBr_3$ interlayer coupling to create effective chiral interlayer fields in the $CrBr_3$ film. Eventually, layered FM and AFM skyrmion lattices emerge in the $CrBr_3$ film, inheriting the moiré periodicity of the interfacial one. Adjacent layered skyrmion lattices exhibit identical (opposite) spin textures when the $CrBr_3$ interlaying coupling between them is FM (AFM). We generated 2 and 3 layered moiré skyrmion lattices in the $(CrBr_3)_2/(CrI_3)_2$ and $(CrBr_3)_3/(CrI_3)_2$ heterostructures, respectively. Meanwhile, attempting to generate a larger number of $CrBr_3$ layered skyrmion lattices by further increasing the $CrBr_3$ film's thickness was found to be impractical.

Finally, we explored the possibility of creating $CrI_3$ skyrmion lattices in heterostructures where the $CrBr_3$ film is considerably thicker than the $CrI_3$ substrate. We successfully generated a single skyrmion lattice in the $(CrBr_3)_3/CrI_3$ heterostructure. However, the formation of layered $CrI_3$ skyrmion lattices proved to be challenging.

## 3. Results and Discussion

### 3.1. Theoretical Models

Following recent experimental works[41,42], we consider a Heisenberg model with intrinsic next NN DMI (IDMI) to describe the intralayer interactions in $CrI_3$ and $CrBr_3$. In the $CrBr_3/CrI_3$ hetero-bilayer, the $CrI_3$ and $CrBr_3$ constitute the bottom ($l=1$) and top ($l=2$) layers, respectively. The top layer is twisted by an angle $\theta$ relative to the bottom layer. The Heisenberg-IDMI Hamiltonian of the hetero-bilayer can be expressed as

$$\mathcal{H}_{BL} = \mathcal{H}_1 + \mathcal{H}_2 + \mathcal{H}_{12}$$

(1)

Here, $\mathcal{H}_1$ ($\mathcal{H}_2$) represents the $CrI_3$ ($CrBr_3$) monolayer Hamiltonian, while $\mathcal{H}_{12}$ accounts for the moiré interlayer exchange interaction. Their explicit forms are given by,

$$\mathcal{H}_l = -\sum_{i,j} J_l\, \mathbf{S}_{li}\cdot\mathbf{S}_{lj} - \sum_i \mathcal{A}_l\, (\mathbf{S}_{li}\cdot\hat{\mathbf{z}})^2 + \sum_{m,n} D_l\, \hat{\mathbf{D}}_{mn}\cdot\mathbf{S}_{lm}\times\mathbf{S}_{ln} + d\sum_{i,j} \hat{\mathbf{d}}_{ij}\cdot\mathbf{S}_{li}\times\mathbf{S}_{lj}$$
$$- \sum_i \mathbf{B}\cdot\mathbf{S}_{li}$$

(2a)

$$\mathcal{H}_{12} = -\sum_{i,j} J_\perp(\mathbf{r}_{ij})\mathbf{S}_{1i}\cdot\mathbf{S}_{2j}$$

(2b)

The vector $\mathbf{S}_{li}$ denotes the spin on a site $i$ in layer $l$ ($l=1,2$). The first term in $\mathcal{H}_l$ accounts for the NN intralayer exchange with strength $J_l$ in layer $l$. The second term is the single-ion uniaxial anisotropy with strength $\mathcal{A}_l$. The third term is the nonchiral IDMI with strength $D_l$. The fourth term accounts for the interfacial chiral NN DMI induced by symmetry breaking[53]. Schematic illustrations of the chiral and nonchiral DM interactions can be found in the supplementary material of Reference[30]. The last term in $\mathcal{H}_l$ accounts for the Zeeman coupling due to an external magnetic field. We adopt experimental values for the parameters $J_l$, $\mathcal{A}_l$, and $D_l$ (Table 1).

$\mathcal{H}_{12}$ represents the moiré interlayer exchange Hamiltonian, where $J_\perp(\boldsymbol{r}_{ij})$ is the distance-dependent interlayer coupling between spins $\mathbf{S}_{1i}$ and $\mathbf{S}_{2j}$. Reference[48] used DFT to calculate the stacking-dependent interlayer exchange energy in $(CrBr_3/CrI_3)$ hetero-bilayer, revealing an AFM interlayer coupling at the local $R_{33}$, $R_{03}$, and $R_{30}$ regions of the moiré supercell. Meanwhile, the interlayer coupling is found FM elsewhere in the moiré supercell. We derive the distance-dependent interlayer coupling from the DFT results utilizing the theoretical method established in our previous work[30], where interested readers can find an in-depth explanation of the approach. This effort allows us to simulate the full spin dynamics in the hetero-bilayers.

The Hamiltonians for the remaining heterostructures can be constructed in a similar manner. Consider, for example, the four-layer (4L) heterostructure $(CrBr_3)_2/(CrI_3)_2$. Here, the $CrBr_3$ homo-bilayer is stacked in a homogeneous AB (R-type) configuration, ensuring a uniform FM (AFM) interlayer coupling within the homo-bilayer[34,46]. The interlayer exchange coefficients in the R-type and AB stacked $CrBr_3$ homo-bilayers are adopted from Reference[34]. The pristine $CrBr_3$ bilayer is twisted by an angle $\theta$ relative to the $CrI_3$ bilayer substrate. The substrate is stacked in an AB configuration, generating a uniform FM interlayer coupling in $CrI_3$. The corresponding exchange coefficient is adopted from Reference[43]. Labeling the layers by $l = 1, 2, 3$, and 4 from bottom to top, the 4L Hamiltonian reads,

$$\mathcal{H}_{4L} = \mathcal{H}_1 + \mathcal{H}_2 + \mathcal{H}_3 + \mathcal{H}_4 + \mathcal{H}_{12} + \mathcal{H}_{23} + \mathcal{H}_{34}$$

(3)

The Hamiltonians $\mathcal{H}_2$ and $\mathcal{H}_3$ for the interfacial layers can be deduced directly from Equation 2a. Similarly, Equation 2a yields the Hamiltonians $\mathcal{H}_1$ (bottom $CrI_3$ layer) and $\mathcal{H}_4$ (top $CrBr_3$ layer) when the NN DMI is set to zero ($d = 0$). Note that the NN DMI is absent in these Hamiltonians because the inversion symmetry is preserved in the top and bottom layers. The Hamiltonian $\mathcal{H}_{23}$ accounts for the moiré interlayer exchange at the interface, treated as in the bilayer case. The terms $\mathcal{H}_{12}$ and $\mathcal{H}_{34}$ represent the uniform interlayer couplings in the pristine $CrI_3$ and $CrBr_3$ bilayers, respectively. These can be expressed as $\mathcal{H}_{12} = -J_{12} \sum_{i,j} \mathbf{S}_{1i} \cdot \mathbf{S}_{2j}$ and $\mathcal{H}_{34} = -J_{34} \sum_{i,j} \mathbf{S}_{3i} \cdot \mathbf{S}_{4j}$, where $J_{12}$ and $J_{34}$ are the interlayer exchange coefficients within the $CrI_3$ and $CrBr_3$ bilayers, respectively.

| Heisenberg-DM model | Parameters |
|---|---|
| $CrI_3$ | $J_1 = 2.13\ meV, D_1 = 0.19\ meV, \mathcal{A}_1 = 0.22\ meV.$ |
| $CrBr_3$ | $J_2 = 1.48\ meV, D_2 = 0.22\ meV, \mathcal{A}_2 = 0.02\ meV.$ |

Table 1: Experimental parameters for the Heisenberg-IDMI models of $CrI_3$ [41] and $CrBr_3$ [42]. The parameters $J$, $D$, and $\mathcal{A}$ stand for the nearest-neighbor intralayer exchange, next NN IDMI, and the single-ion uniaxial anisotropy, respectively.

## 3.2. Computational Details

The heterostructure Hamiltonian is employed in the atomistic sLLG equation,

$$\frac{\partial \mathbf{S}_{li}}{\partial t} = -\frac{\gamma}{1+\lambda^2}\left[\mathbf{S}_{li} \times \mathbf{H}_{li}^{eff} + \lambda \mathbf{S}_{li} \times \left(\mathbf{S}_{li} \times \mathbf{H}_{li}^{eff}\right)\right]$$

(4)

where $\gamma$ and $\lambda$ are the gyromagnetic ratio and Gilbert damping, respectively. The effective field $\mathbf{H}_{li}^{eff}$ acting on spin $\mathbf{S}_{li}$ is given by

$$\mathbf{H}_{li}^{eff} = -\frac{1}{\mu_s}\frac{\partial \mathcal{H}}{\partial \mathbf{S}_{li}} + \mathbf{H}_{li}^{th}$$

(5)

In Equation 5, $\mathcal{H}$ stands for the heterostructure Hamiltonian, while $\mathbf{H}_{li}^{th}$ is the effective thermal field. We use the Vampire software[57] to solve the sLLG equations for all the spins in the system and eventually simulate the spin dynamics in the entire heterostructure. $\mathbf{H}_{li}^{th}$ is included in Vampire using the Langevin Dynamics method[58]. The input files for the Vampire software are prepared using a Mathematica[59] code developed by the authors.

The sLLG simulations are initiated at a high temperature ($50\ K$) with random initial spin configurations. Each simulation was repeated employing at least three distinct random spin configurations to test the dependence of the results on the initial state. The moiré heterostructures are cooled gradually using $1 \times 10^{-16} s$ for the time step, a $1\ ns$ cooling time, and a total of $3 \times 10^7$ time steps. The sLLG simulations yield the time evolution of the spin configurations in all layers at temperatures down to the ground state near $0\ K$. To prevent the formation of spin spirals, an external magnetic field ($50 \sim 100\ mT$) is applied during cooling. This field is deactivated after cooling, with no impact on the stability of the achieved ground state. Simulations are conducted on both single and multiple moiré supercell samples to validate the spatial periodicity of the topological ground states.

### 3.3. Skyrmion lattices in moiré $CrBr_3/CrI_3$ hetero-bilayers

We carried out sLLG simulations on moiré $CrBr_3/CrI_3$ hetero-bilayers with commensurate twist angles between $1.02°$ and $3.15°$. In the absence of DFT and experimental estimates for the strength of NN DMI in these hetero-bilayers, we treated the NN DMI as a parameter, varied between 0 and $0.3\ meV$ in steps of $0.01\ meV$. The assumption of a sizeable NN DMI in the hetero-bilayers is justifiable (due to the interface) and supported by recent experimental evidence[53]. Moreover, the twisting and external factors (e.g., an applied electric field[60]) might enhance the NN DMI strength.

The interlayer interaction in the moiré supercell of a twisted $CrBr_3/CrI_3$ hetero-bilayer is AFM in the local $R_{33}$, $R_{03}$ (equivalently $R_{63}$), and $R_{30}$ (equivalently $R_{36}$) stacking regions, and FM elsewhere[48]. Below a critical value of the twist angle, the interlayer interaction dominates the intralayer exchange, inducing AFM domains at the local AFM regions, similar to moiré $CrI_3$ homo-bilayers[30–32]. When the chiral NN DMI is weak, the AFM domains can be topological or trivial, depending on the random initial spin configuration[30,34]. Here, the vorticity and helicity of the AFM-FM domain wall act as degrees of freedom, resulting in trivial spin textures or multi-flavored TSTs in an unpredictable manner[30,34]. On the other hand, a sizeable NN DMI (expected in hetero-bilayers[61]) locks the degrees of freedom, leading to the deterministic creation of Néel-type skyrmions at the local AFM regions. Figures 1a-d elaborates this process in a moiré $CrBr_3/CrI_3$ hetero-bilayer with a small twist angle ($\theta = 1.02°$) and a sizeable NN DMI ($d = 0.3\ meV$).

The simulation starts at $T = 50\ K$, assuming random initial spins in both layers (Figure 1a) and applying a $50\ mT$ field during the cooling process. At relatively high temperatures, thermal fluctuations dominate the magnetic interactions, hindering the formation of magnetic order in the hetero-bilayer. As the temperature decreases and thermal fluctuations diminish, magnetic domains begin to form in both layers (Figure 1b). With further cooling, coexisting FM and AFM domains appear in the $CrBr_3$ layer (Figure 1c), while the bottom $CrI_3$ becomes FM. At this stage, the weaker intralayer exchange in $CrBr_3$ compared to $CrI_3$ (Table 1) makes it energetically favorable for the spin textures to form exclusively in the $CrBr_3$ layer. Nevertheless, the AFM domains at these temperatures do not display an ideal topological texture (Figure 1c, e). Upon further cooling, the influence of the NN DMI becomes evident in shaping the AFM domains' spin morphology and creating $CrBr_3$ Néel-type skyrmions near $0\ K$ (Figure 1d, f). The magnetic field can now be removed without compromising the stability of the skyrmions.

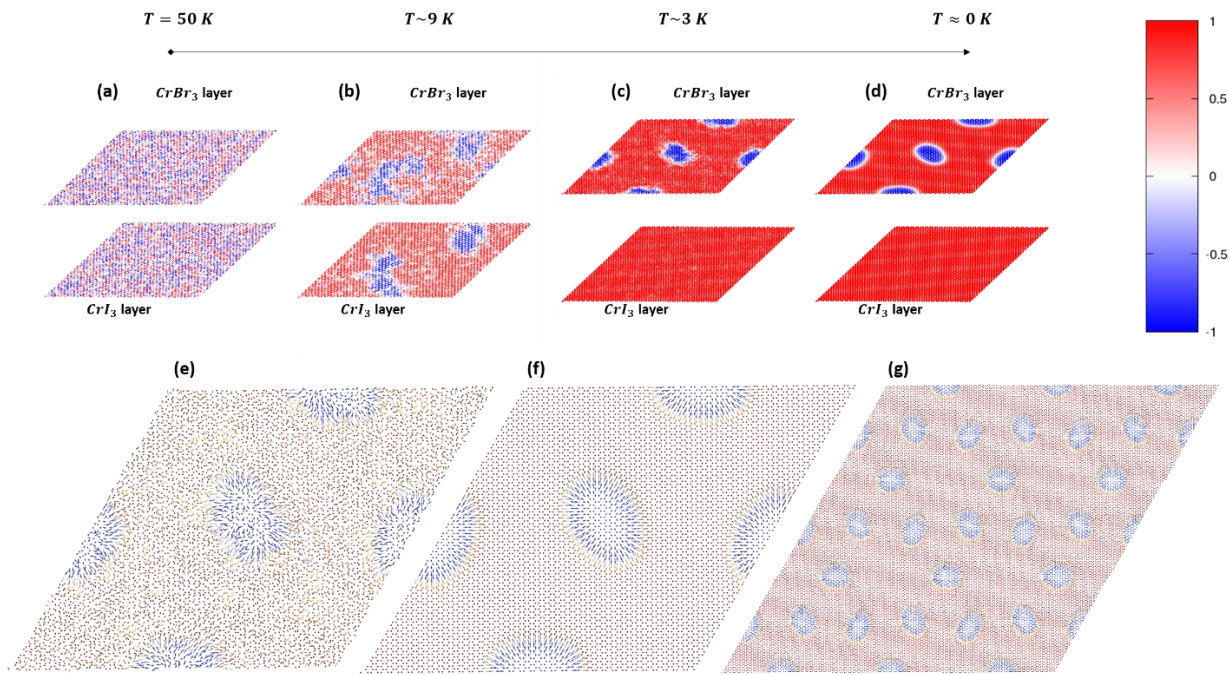

**Fig. 1: Controlled nucleation of Néel-type skyrmion lattices in moiré $CrBr_3/CrI_3$ hetero-bilayers**. Cooling the $CrBr_3/CrI_3$ hetero-bilayer from high temperatures and random spins results in skyrmion nucleation within the $CrBr_3$ top layer (**a-d**). The weaker intralayer exchange in $CrBr_3$ compared to $CrI_3$ allows for controlled skyrmion formation in the top layer, while the $CrI_3$ remains in a ferromagnetic state. While cooling, AFM domains forms in $CrBr_3$ at regions with local AFM interlayer coupling in the moiré supercell. The chiral Dzyaloshinskii-Moriya interaction (DMI) gradually shapes the AFM domains' spin texture, achieving Néel-type skyrmions in the ground state (**e, f**). In a multi-moiré simulation, the controlled $CrBr_3$ skyrmion nucleation creates an ordered skyrmion crystal (**g**). Results are shown for a hetero-bilayer with 1.02° twist angle and 0.3 $meV$ NN DMI. The color scale represents the magnitude of the out-of-plane spin components, ranging from -1 (blue) to 1 (red).

The profile of the interlayer exchange in the moiré supercell[48] determines the position, size, and elliptic form of the $CrBr_3$ skyrmions. Because the moiré skyrmions are relatively well-spaced, interactions between skyrmions are minimal, preserving their Néel-type spin structure. In samples with multiple supercells, this scenario repeats in every moiré cell, resulting in an ordered moiré skyrmion lattice, as shown in Figure 1g. This level of control over 2D moiré skyrmions provides a significant advantage over DMI-induced skyrmions in conventional materials, which form in disordered configurations and inconsistent morphologies.

As previously discussed, the modulated interlayer interaction can generate AFM domains for small twist angles even without NN DMI. However, to produce an ordered skyrmion lattice, the NN DMI must be present and strong enough to create three Néel-type skyrmions per moiré supercell (a phase we denote as the "3 skyrmions phase"). To determine the threshold DMI necessary for the formation of the 3 skyrmions phase, we performed sLLG simulations at commensurate angles within the range of $1.02° \lesssim \theta \lesssim 3.15°$, while systematically varying the DMI from 0 to $0.3\ meV$ in increments of $0.01\ meV$ (Figure 2). To validate the emergence of the 3 skyrmions phase, we repeated each sLLG simulation five times, starting from different initial spin configurations. The analysis reveals an intriguing trend, showcasing a notable decrease in the threshold DMI as the twist angle is reduced. As an example, Figure 2 demonstrates the formation of the 3 skyrmions phase at a remarkably low threshold DMI of approximately $0.08\ meV$ for $\theta \approx 1.02°$. Conversely, a relatively strong NN DMI ($d = 0.3\ meV$) generates moiré skyrmion lattices up to a twist angle $\theta \approx 2.64°$.

The main focus of this study is the 3 skyrmions phase, which produces moiré skyrmion lattices. However, we briefly examined the behavior below the threshold DMI for completeness. In this case, the DMI is insufficient to form three Néel-type skyrmions. As a result, the spin helicity and vorticity within some AFM domains become active degrees of freedom, leading to either trivial or topological spin textures depending on the initial random spin configuration[30,34]. Consequently, below the threshold DMI, one expects the emergence of additional phases that complement the 3 skyrmions phase, specifically the $n = 2, 1$, and 0 skyrmion phases. In these phases, there are $n$ Néel-type skyrmions that are relatively robust to variations in the initial spin configuration, while the remaining $3 - n$ AFM domains can be either trivial or topological depending on their initial state. We have partially succeeded in detecting these phases below the threshold DMI, as depicted in Figure 2 and Supplementary Figures 1-3. However, it is important to note that these phases do not result in a periodic skyrmion lattice and are not directly relevant to the main objective of this study.

Figure 2 further illustrates the critical twist angles for the formation of AFM domains. When the NN DMI is negligibly small, the magnetic phases with AFM domains start to appear below a critical angle $\theta_c \approx 2.15°$ (Figure 2), due to the dominant interlayer interaction. Meanwhile, a sizeable DMI assists the interlayer coupling in dominating the intralayer exchange, raising $\theta_c$ to approximately $2.85°$ at $d = 0.3\ meV$ (Figure 2). Note that these values for $\theta_c$ are inferior to the one reported in $CrI_3$ homo-bilayers[30–32] due to the weak intralayer exchange in $CrBr_3$ compared to $CrI_3$. Above $\theta_c$, the hetero-bilayer is essentially FM (see the FM phase in Figure 2), with slightly tilted spins at the local AFM stacking regions.

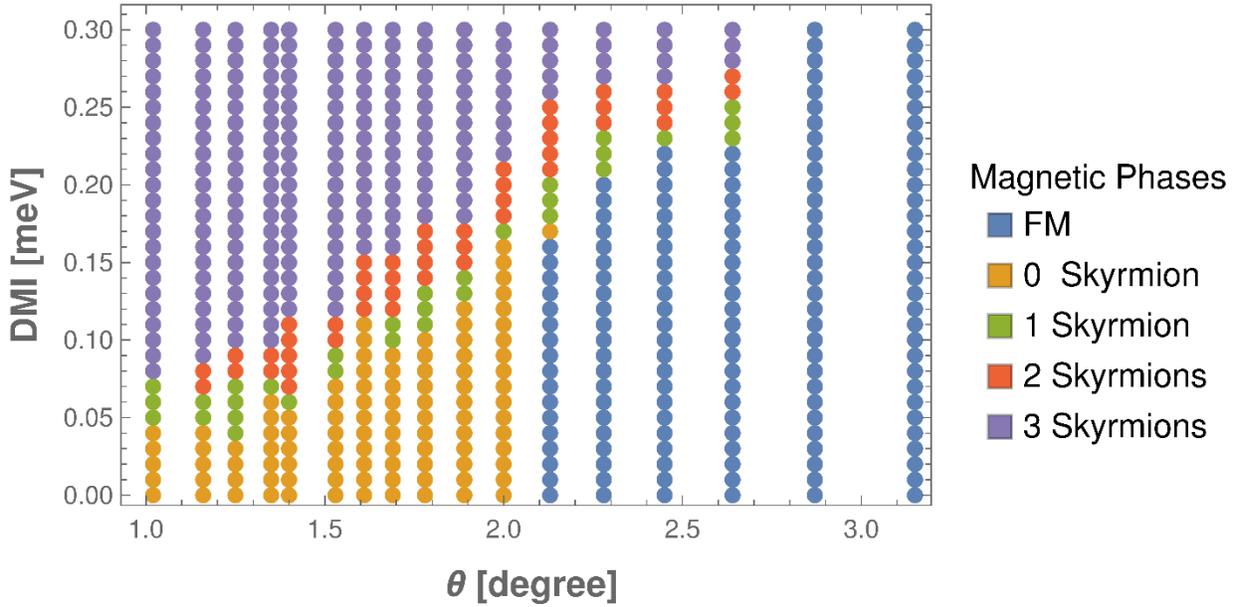

**Fig. 2: Magnetic phases in moiré $CrBr_3/CrI_3$ hetero-bilayers**. The figure illustrates the magnetic phases in a moiré $CrBr_3/CrI_3$ hetero-bilayer, considering commensurate twist angles in the range $1.02° \lesssim \theta \lesssim 3.15°$ and chiral Dzyaloshinskii-Moriya interaction (DMI) between 0 and $0.3\ meV$. The main result is the 3 skyrmion phase, which generates a moiré skyrmion lattice. This phase is achievable above a twist-dependent threshold DMI. The figure reveals a notable decrease in the threshold DMI as the twist angle is reduced. Below its threshold value, the DMI is insufficient to form three Néel-type skyrmions per moiré supercell. With limited success, we detected $n = 2, 1$, and 0 skyrmion phases below the threshold DMI, which complement the 3 skyrmions phase (see text for details).

In closing this paragraph, we briefly address the manipulation of the skyrmion lattice by a perpendicular magnetic field $\boldsymbol{B}$ at $0\ K$. When a magnetic field is applied opposite to the skyrmions' spins, the skyrmions shrink and ultimately disappear. Conversely, a magnetic field parallel to the spins inflates the skyrmions and eventually induces a global magnetization reversal in the hetero-bilayer. By gradually reducing the magnetic field to $0\ T$ following the magnetization reversal, a skyrmion lattice forms with spins opposite to the initial configuration. This hysteresis behavior is extensively discussed in the Supplementary Information, and a visual demonstration is provided in Supplementary Video 1.

### 3.4. Layered skyrmion lattices in moiré $(CrBr_3)_n/(CrI_3)_{n'}$ heterostructures

In this section, we extend the study beyond hetero-bilayers and investigate the formation of layered moiré skyrmion lattices. Specifically, we demonstrate the emergence of 2 and 3 layered $CrBr_3$ skyrmion lattices in $(CrBr_3)_2/(CrI_3)_2$ and $(CrBr_3)_3/(CrI_3)_2$ moiré heterostructures, respectively.

We first consider the 4L $(CrBr_3)_2/(CrI_3)_2$ heterostructures, derived from a pristine $CrBr_3$ bilayer twisted by an angle $\theta$ relative to a pristine $CrI_3$ bilayer substrate. The substrate is stacked in an FM (i.e., AB) configuration, while the $CrBr_3$ bilayer is stacked in an FM (i.e., AB) or an AFM (i.e., $R$- type) configuration[31,34,45,46]. The magnetic ground state of the 4L heterostructure is determined by several competing interactions. These include the intralayer interactions in all layers, the moiré-modulated interlayer field at the interface between $CrI_3$ and $CrBr_3$, and the uniform FM or AFM interlayer couplings in the ultrathin $CrBr_3$ and $CrI_3$ films. However, the relatively strong exchange interactions in the $CrI_3$ substrate ensure that any emerging spin texture remains confined to the $CrBr_3$ ultrathin film.

The high computational cost to simulate the 4L heterostructure renders a comprehensive study across the ranges $1.02° \lesssim \theta \lesssim 3.15°$ and $0 \leq d \leq 0.3$ unfeasible. Instead, we concentrated our efforts on the angle $\theta = 1.02°$ and estimated the threshold NN DMI needed to produce layered skyrmion lattices. These thresholds were found to be $0.18\ meV$ and $0.16\ meV$ for 4L heterostructures with FM and AFM $CrBr_3$ films, respectively. Based on the observed trend in the hetero-bilayer case, further reducing the twist angle in an experiment allows for the realization of layered skyrmion lattices at lower DMI values. Figure 3 illustrates our findings for $\theta = 1.02°$ and the corresponding threshold DMI values.

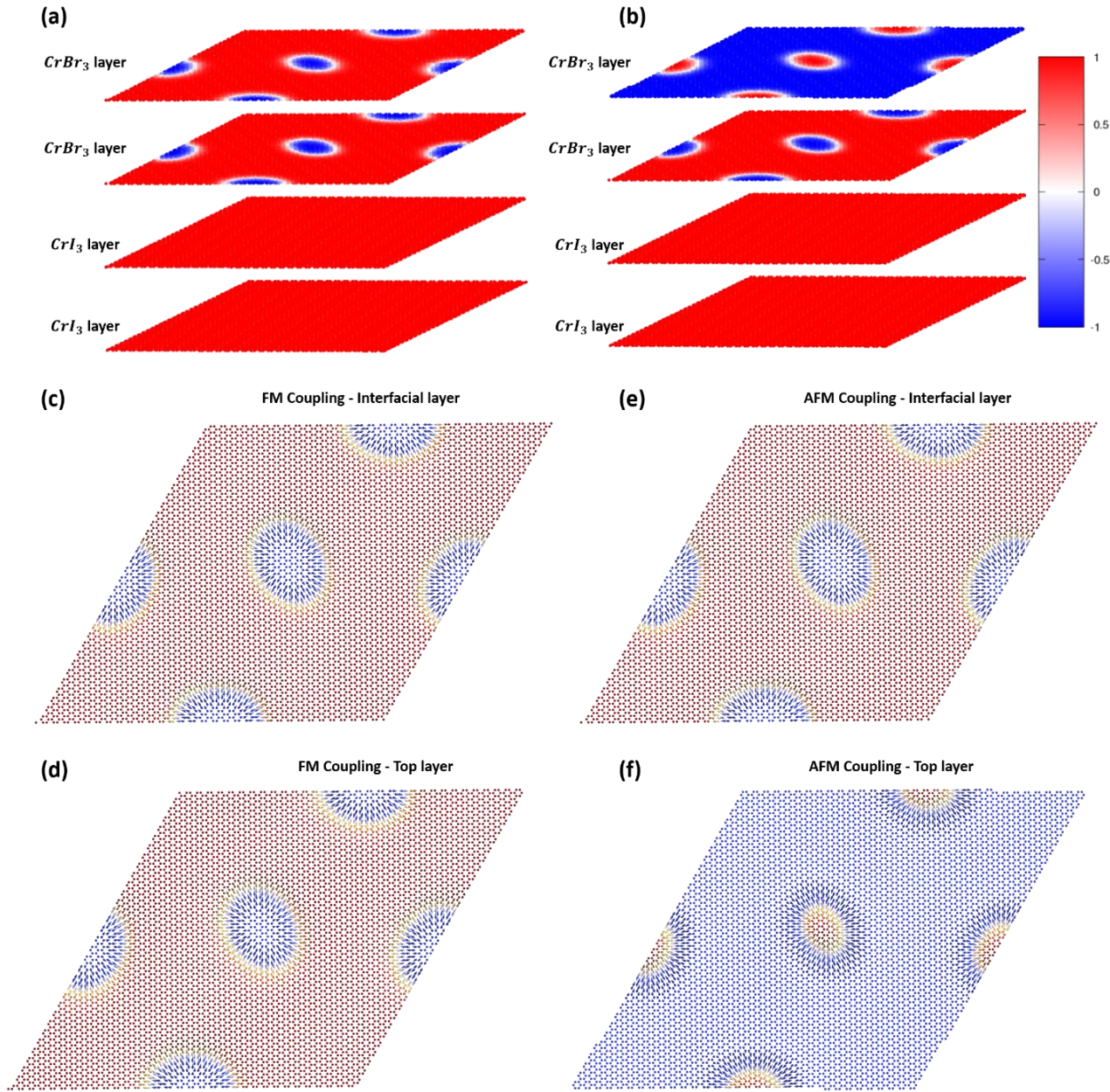

**Fig. 3: Layered skyrmion lattices in moiré four-layer magnetic heterostructures**. (**a**) $(CrBr_3)_2/(CrI_3)_2$ heterostructures formed from an AB-stacked $CrBr_3$ bilayer, twisted by an angle $\theta = 1.02°$ relative to an AB-stacked $CrI_3$ substrate. Néel-type moiré skyrmions are nucleated in the interfacial $CrBr_3$ layer due to the moiré interlayer coupling and interfacial DMI (**a**, **c**). The uniform FM interlayer coupling and skyrmionic spin textures endow the top $CrBr_3$ layer with an effective chiral interlayer field, imprinting skyrmions in the top layer (**a**, **d**). Consequently, the skyrmions in the top layer inherit their spatial arrangement and chirality from those in the middle layer (**a, c, d**). The heterostructure in (**b**) is similar to that in (**a**), but the $(CrBr_3)_2$ bilayer is stacked in an R-type (AFM) configuration, resulting

in skyrmions in the top layer with opposite chirality compared to the interfacial $CrBr_3$ layer (b, e, f). The results for the heterostructures in (**a**) and (**b**) are obtained using the minimal DMI values required to generate the 3 Skyrmion phase, namely $0.18\ meV$ and $0.16\ meV$, respectively. Increasing the sample's size generates layered 2D skyrmion lattices. The color scale represents the out-of-plane spin components, ranging from -1 (blue) to 1 (red).

Figure 3a illustrates the FM $CrBr_3$ case. In the ground state, the moiré interlayer (assisted by the NN DMI) stabilizes three Néel-type skyrmions in the interfacial $CrBr_3$ layer (Figure 3a, c). The uniform FM interlayer coupling in $CrBr_3$ combined with the skyrmionic spin textures in the interfacial $CrBr_3$ layer provides an effective chiral interlayer field to the top $CrBr_3$ layer, stabilizing three Néel-type skyrmions in this DMI-free layer (Figure 3a, d). Due to the FM interlayer coupling, the skyrmions in the top and interfacial $CrBr_3$ layers are concentric with matching chirality (Figure 3a, c, d). In a multi-moiré sample, the vertical imprinting of the moiré skyrmions is replicated in all moiré supercells, resulting in layered 2D $CrBr_3$ skyrmion lattices with matching spin textures. Figure 3b corresponds to the AFM $CrBr_3$ case. As before, the 3 skyrmion phase is achieved in the interfacial $CrBr_3$ layer due to the moiré interlayer coupling and NN DMI (Figure 3b, e). In the present case, due to the AFM $CrBr_3$ interlayer coupling, the effective interlayer field on the top $CrBr_3$ layer has opposite chirality compared to the interfacial layer's skyrmions. As a result, the effective chiral field creates skyrmions in the top layer with opposite spins compared to the interfacial layer (Figure 3b, e, f). Expanding the sample size to include multiple moiré supercells generates layered 2D $CrBr_3$ skyrmion lattices with opposite spin alignments in the $CrBr_3$ bilayer.

Next, we consider the 5L $(CrBr_3)_3/(CrI_3)_2$ heterostructure, composed of a pristine $CrBr_3$ trilayer twisted by an angle $\theta$ relative to a pristine $CrI_3$ bilayer substrate. The $CrI_3$ substrate remains in an FM configuration, while the interlayer $CrBr_3$ coupling can be FM or AFM between adjacent layers.

In the 5L heterostructure case, creating layered skyrmion lattices becomes challenging within the previously adopted ranges for $\theta$ and $d$. To avoid using unreasonably high values of the NN DMI, we reduce the twist angle below $1.02°$ and consider a heterostructure with $\theta = 0.91°$ and $d = 0.3\ meV$. The results from 12 simulations (with different initial random spins) indicated that the formation of skyrmions within the $CrI_3$ bilayer substrate is energetically prohibited, while robust layered skyrmion lattices are found to emerge in the $CrBr_3$ trilayer. Analogous to previous heter-ostructures, the moiré field and the NN DMI collaborate to generate moiré skyrmions in the interfacial $CrBr_3$ layer. These are then imprinted vertically into the remaining $CrBr_3$ layers via the chiral interlayer fields (Figure 4). By combining FM and AFM stacking in the $CrBr_3$ trilayer, we successfully designed 4 distinct layered skyrmion lattices, as depicted in Figure 4.

In our 5L simulations, we selected $d = 0.3\ meV$ to facilitate the formation of layered skyrmion lattices, even though this value may surpass the threshold DMI. As previously emphasized, reducing the twist angle further will decrease the threshold DMI required for achieving these 3 layered skyrmion lattices. Importantly, because of the 2L and 3L thicknesses of the $CrI_3$ and $CrBr_3$ films in the heterostructure, the twist angle can be reduced to tiny values without compromising the structural stability of the interface. This is supported by a recent experimental study[29], demonstrating robust structural stability of the moiré interface between slightly twisted ultrathin $CrI_3$ films with twist angles smaller than ~0.5°.

Our examination of moiré $(CrBr_3)_n/(CrI_3)_{n'}$ heterostructures with $n \geq 4$ revealed that generating 4 or more layered skyrmion lattices is not feasible for the selected materials. Additionally, we briefly explored the potential of generating moiré skyrmions in the $CrI_3$ substrate, using the $(CrBr_3)_3/CrI_3$ heterostructure as an initial trial. This heterostructure comprises a pristine FM-stacked $CrBr_3$ trilayer twisted atop a single $CrI_3$ monolayer. Indeed, at large moiré periodicity and sizeable NN DMI, the formation of a single $CrI_3$ skyrmion lattice was found to be energetically favorable in this heterostructure (Supplementary Figure 4). However, our subsequent attempts to generate layered $CrI_3$ skyrmion lattice in $(CrBr_3)_4/(CrI_3)_2$ and $(CrBr_3)_5/(CrI_3)_2$ moiré heterostructures were unsuccess. Specifically, simulations of these heterostructures at $\theta = 0.91°$ and $d = 0.3\ meV$ resulted in a layered FM ground state.

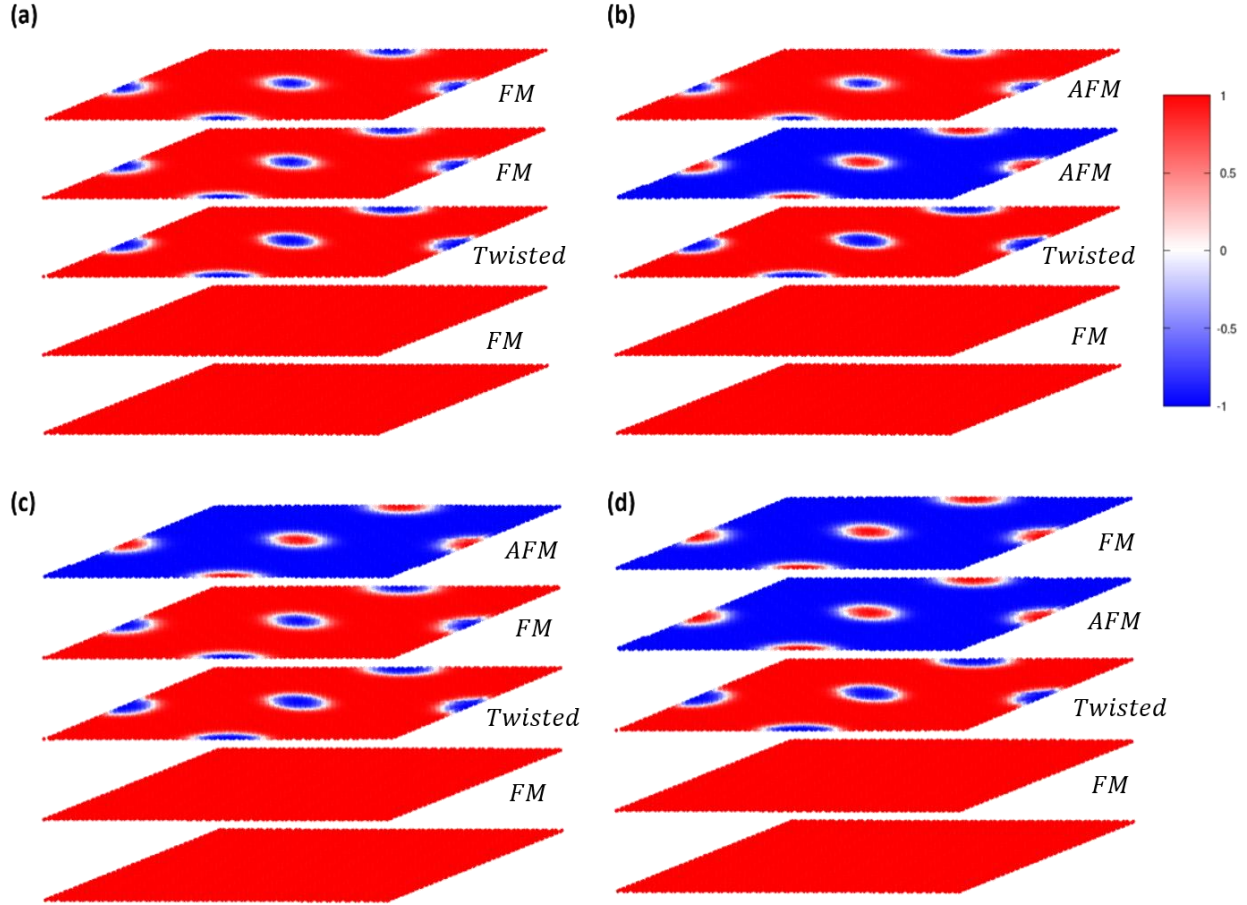

**Fig. 4: Layered skyrmion lattices in moiré five-layer magnetic heterostructure**. Layered skyrmion lattices in $(CrBr_3)_3/(CrI_3)_2$ heterostructure, composed of a $CrBr_3$ trilayer (top three layers) twisted by an angle $\theta = 0.91°$ relative to a $CrI_3$ bilayer substrate (the two bottom layers). The $CrI_3$ substrate remains in a ferromagnetic (FM) configuration, while the interlayer $CrBr_3$ coupling can be FM or antiferromagnetic (AFM) between adjacent layers, as illustrated in Figures (**a**)-(**c**). The value of the Dzyaloshinskii-Moriya interaction (DMI) is set to $0.3\ meV$ in the interfacial layers. The moiré field and the NN DMI generate Néel-type moiré skyrmions in the interfacial $CrBr_3$ layer. These are then imprinted vertically into the remaining $CrBr_3$ layers via emergent chiral interlayer fields. The moiré skyrmions in adjacent layers display identical (opposite) spins when the corresponding pristine interlayer coupling is FM (AFM). The color scale represents the magnitude of the out-of-plane spin components, ranging from -1 (blue) to 1 (red).

## 4. Conclusions

Skyrmions can nucleate in conventional magnetic thin films and multilayers due to bulk or interfacial DMI[62–64]. However, implementing DMI skyrmions in future spintronic and magnonic devices is challenged by their field-assisted stabilization and the lack of control over their spatial distribution[54–56]. Moiré skyrmions could address these challenges, as their nucleation process fundamentally differs from conventional DMI skyrmions.

Below a critical twist angle, spin textures form in moiré magnets at specific locations in the moiré supercell where the interlayer coupling is locally AFM. A sizeable NN DMI shapes the spin textures into Néel-type skyrmions, moderately elevating the critical twist angle value. The NN DMI, however, does not affect the skyrmions' spatial distribution, determined exclusively by the interlayer coupling. Therefore, moiré magnets hold promises for the controlled nucleation of skyrmions at predefined locations.

Nevertheless, skyrmions in single-material moiré magnetic structures can form in any layer, hindering the formation of periodic 2D arrangements of skyrmions. In this regard, moiré magnetic heterostructures offer valuable advantages over their single-material counterparts. Interfaces in these heterostructures induce sizeable NN DMI[61], while the variation in magnetic properties across the heterostructure confines skyrmions to specific layers. Accordingly, these heterostructures might serve as good candidates to realize 2D skyrmion lattices.

However, moiré magnetic heterostructures remain largely unexplored, lacking experimental studies and advanced modeling that can account for the complex interplay of their competing interactions. The current study highlighted their potential to create skyrmion architectures that might not be feasible in conventional magnetic films and multilayers. We chose $CrI_3$ and $CrBr_3$ to demonstrate our results, primarily due to the wealth of available data on their stacking-dependent magnetism and intralayer interactions. However, it is essential to note that the implications of our findings go beyond these specific materials, which only serve as illustrative prototypes.

We simulated the skyrmion phases in moiré $(CrBr_3)_n/(CrI_3)_{n'}$ heterostructures, including all spin degrees of freedom, intralayer interactions, thermal effects, and interlayer couplings. The proposed heterostructures are derived from a pristine $CrBr_3$ ultrathin film ($n$ layers), twisted relative to a pristine $CrI_3$ ultrathin film ($n'$ layers). The intralayer interactions in $CrBr_3$ and $CrI_3$ are adopted from experimental models with spin-orbit couplings (next NN DMI). The thermal effects are involved via the sLLG equations, and the interlayer interactions are modeled using available DFT data for the stacking-dependent magnetism in $CrBr_3/CrI_3$, $CrBr_3/CrBr_3$, and $CrI_3/CrI_3$ bilayers.

For appropriate values of the substrate thickness, twist angles, and interfacial NN DMI, layered $CrBr_3$ moiré skyrmion lattices can form in the proposed heterostructure. A moiré skyrmion lattice forms at the $CrBr_3$ interfacial layer, and is subsequently projected to the remaining layers of the $CrBr_3$ film. The spin orientation of the projected skyrmions can be identical or opposite to that of the interfacial moiré skyrmions, depending on the sign of the pristine interlayer coupling in the $CrBr_3$ film.

Despite the modest moiré interlayer interaction in $CrBr_3/CrI_3$, we generated three $CrBr_3$ layered skyrmion lattices in $(CrBr_3)_3/(CrI_3)_2$. A recent theoretical study[65] identified several magnetic hetero-bilayers with more substantial and diverse moiré interlayer coupling. These materials hold promise for future research and may outperform the $(CrBr_3)_n/(CrI_3)_{n'}$ heterostructures in generating layered skyrmion lattices.

Recently, several 2D magnets exhibiting spontaneous magnetization at room temperature and beyond have been reported[25,66–68]. This ongoing effort is crucial for potentially integrating 2D magnets into commercial devices operating at room temperature. Although the layered moiré skyrmion lattices in our work manifest only at low temperatures in the chosen materials, our findings underscore the potential of moiré magnetic heterostructures for engineering novel topological spintronic and magnonic devices. Over recent years, several low-temperature spintronic devices have been proposed following the observation of layered FM and AFM collinear phases in $CrI_3$. These include spin-filter tunnel barriers[69], field-effect transistors[70], nonreciprocal optical devices[71], photovoltaic devices[72], spin current devices[73], and spin valves that can induce conductor-to-insulator transition in graphene[74]. The layered FM and AFM skyrmion lattices revealed in our study extend beyond the standard layered FM and AFM phases observed in pristine $CrI_3$. These topological phases induce chiral magnetic fields and could introduce new intriguing observations and functionality to the previously proposed spintronic devices. With their remarkable periodicity compared to DMI-induced skyrmions, these lattices could offer exciting platforms for the ongoing research on moiré magnons[75–80] and skyrmion-based magnonics.

## 5. Methods

All methods are included in the Results and Discussion section.

## 6. Data availability

The datasets generated during the current study are available from the corresponding author upon reasonable request.

## 7. Code availability

The code developed in this study is available from the corresponding author upon reasonable request.

## Acknowledgments

Part of the numerical calculations was performed using the Phoenix High Performance Computing facility at the American University of the Middle East (AUM), Kuwait.

## Contributions

D.G. planned the research with B.J, developed the theoretical models, and wrote the Mathematica codes. B.J. performed the Vampire simulations, analyzed the results with D.G., prepared the Supplementary Material, and drafted the manuscript. D.G. revised the manuscript and wrote the final version with B.J.

# Supporting Information

# Designing layered 2D skyrmion lattices in moiré magnetic heterostructures

*Bilal Jabakhanji[1] and Doried Ghader*[\*1]*

[1] College of Engineering and Technology, American University of the Middle East, Egaila 54200, Kuwait

## Supporting Information 1. Supplementary Figures.

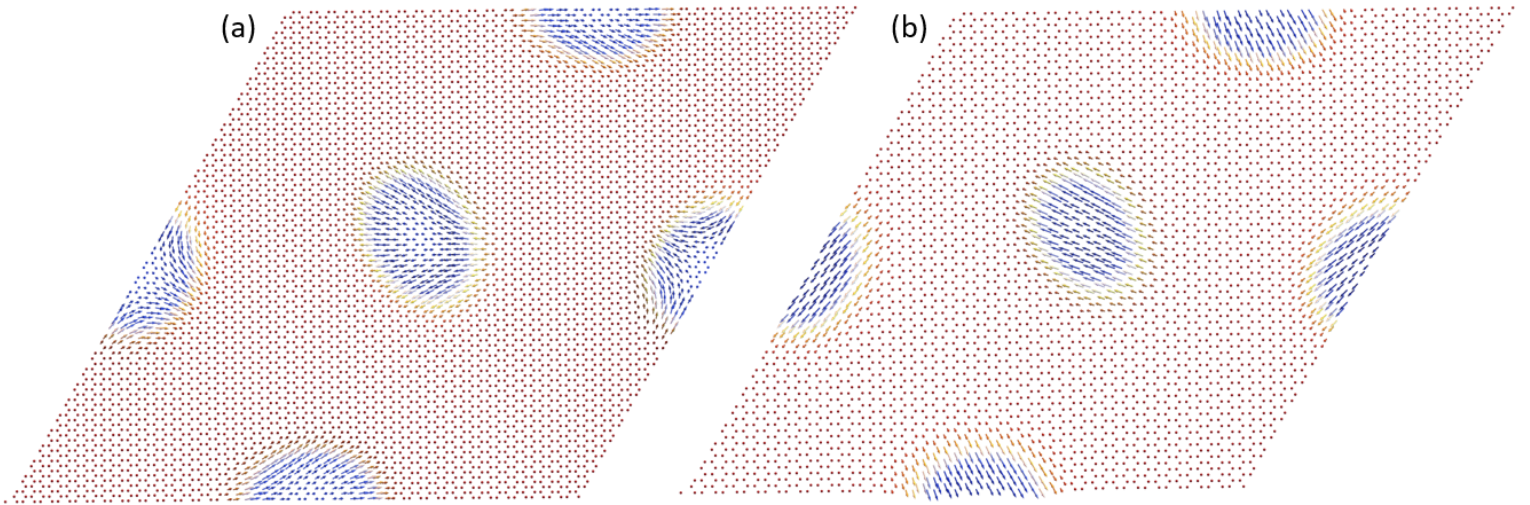

**Figure S1.** Illustration of the "0 skyrmions phase" in the $CrBr_3$ layer of the moiré $CrBr_3/CrI_3$ hetero-bilayers for $\theta \approx 1.02$, $d = 0.03\ meV$ (a) and $\theta \approx 1.25$, $d = 0.01\ meV$ (b). The Néel-type skyrmions are absent in both figures, which is a characteristic of the 0 skyrmions phase. In (a), an anti-skyrmion is observed in the local $R_{03}$ (equivalently $R_{63}$) stacking region of the moiré supercell. The remaining AFM domains are topologically trivial. In (b), all AFM domains are topologically trivial.

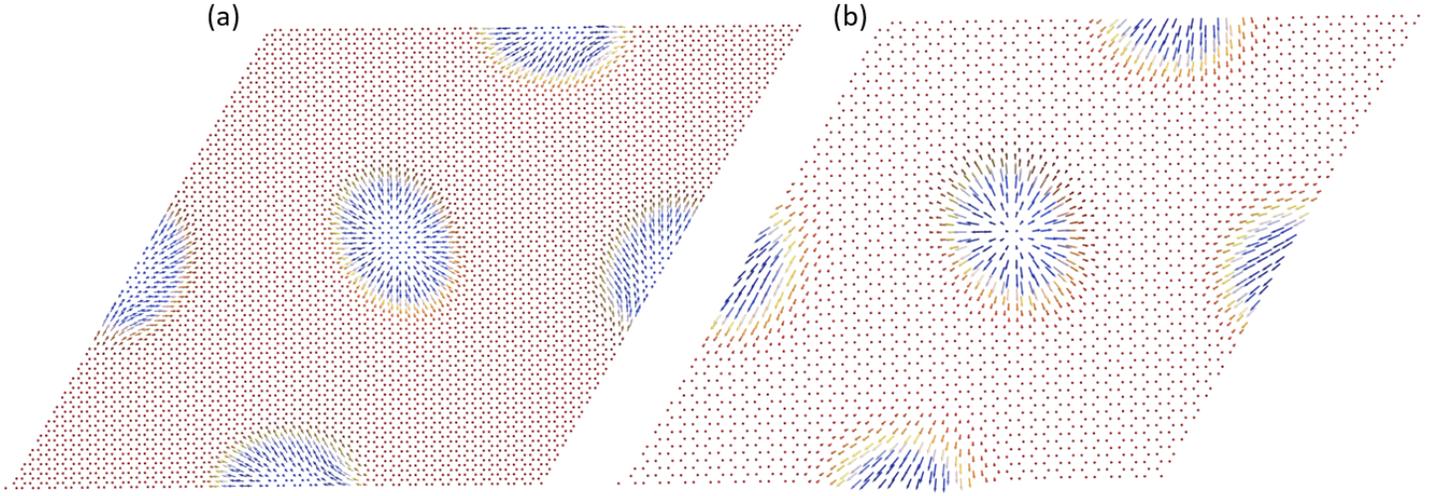

**Figure S2.** Illustration of the "1 skyrmions phase" in the $CrBr_3$ layer of the moiré $CrBr_3/CrI_3$ hetero-bilayers for $\theta \approx 1.02, d = 0.06\ meV$ (a) and $\theta \approx 1.53, d = 0.08\ meV$ (b). In (a) and (b), a single Néel-type skyrmion is observed at the local $R_{33}$ stacking region of the moiré supercell, while the remaining AFM domains are topologically trivial.

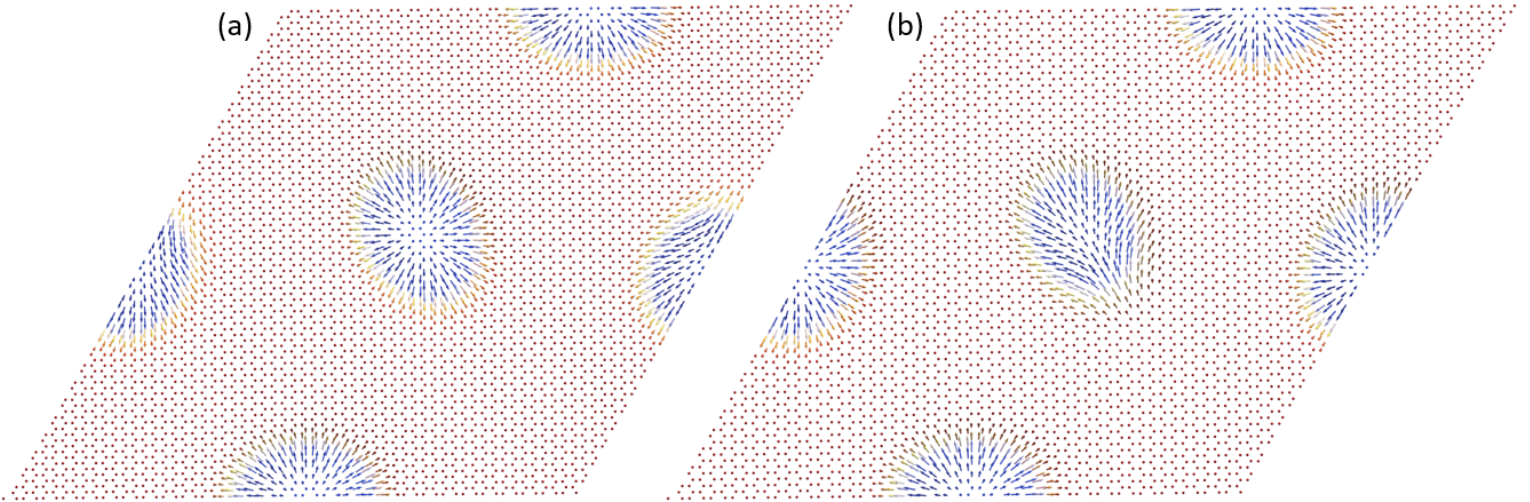

**Figure S3.** Illustration of the "2 skyrmions phase" in the $CrBr_3$ layer of the moiré $CrBr_3/CrI_3$ hetero-bilayers for $\theta \approx 1.16, d = 0.07\ meV$ (a) and $\theta \approx 1.25, d = 0.09\ meV$ (b). In (a), Néel-type skyrmions are observed at the local $R_{33}$ and $R_{30}$ (equivalently $R_{36}$) stacking regions of the moiré supercell. The remaining AFM domain is topologically trivial. In (b), Néel-type skyrmions are observed at the local $R_{30}$ and $R_{03}$ stacking regions, while the remaining AFM domain is topologically trivial.

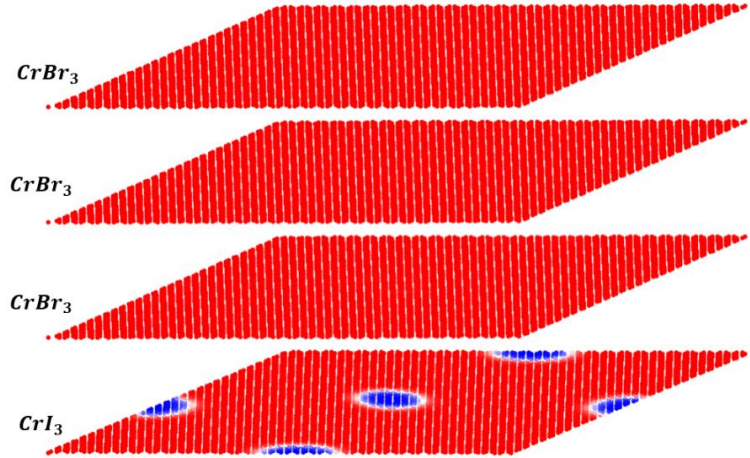

**Figure S4.** The figure illustrates the formation of $CrI_3$ moiré skyrmions in the $(CrBr_3)_3/CrI_3$ heterostructure with $\theta = 1.16°$ and $d = 0.3\ meV$.

## Supporting Information 2. Hysteresis behavior of moiré $CrBr_3/CrI_3$ hetero-bilayers

In the main text, we briefly discussed the manipulation of the moiré skyrmion lattice in $CrBr_3/CrI_3$ hetero-bilayers via magnetic fields at $0\ K$. Here, we provide a detailed description of the hetero-bilayer's hysteresis behavior near $0\ K$ in the skyrmion lattice phase ($\theta = 1.02°$ and $d = 0.3\ meV$) and the FM phase ($\theta = 6°$ and $d = 0.3\ meV$). For ease of reference, we will refer to these hetero-bilayers as a "3 Skyr hetero-bilayer" and an "FM hetero-bilayer". Note that a visual demonstration of the 3 Skyr hetero-bilayer case is available in Supplementary Video 1.

We start by performing sLLG simulations to determine the ground state of the hetero-bilayers. Once the ground state is established, we apply an external magnetic field, varied between $\pm 3\ \hat{z}\ T$ in steps of $1\ mT$, and register the hysteresis behavior. As these hysteresis simulations are conducted near $0\ K$, the sLLG equation simplifies to its LLG version, thereby generating the spin configurations and the average magnetization for each magnetic field value.

Supplementary Figure S5a shows the hysteresis loop for an FM hetero-bilayer with $\theta = 6°$ and $d = 0.3\ meV$. We observe a square-shaped hysteresis loop with abrupt magnetization reversal and large coercive fields ($\sim \pm 2.2\ \hat{\mathbf{z}}\ T$). The persistent magnetic saturation followed by sudden magnetization reversal indicates the lack of stable magnetic domain formation during the hysteresis manipulation of the hetero-bilayer.

The 3 Skyr hetero-bilayer displays a fundamentally different hysteresis behavior, illustrated in Supplementary Figure S5b for $\theta = 1.02°$ and $d = 0.3\ meV$. Labels I-VII highlight the major events during the hysteresis manipulation. From I ($\mathbf{B} = 3\ \hat{\mathbf{z}}\ T$) to II ($\mathbf{B} \approx 1.6\ \hat{\mathbf{z}}\ T$), the hetero-bilayer remains almost FM (Supplementary Figure S5c-I and II) with a saturated average magnetization ($M \approx 1$). From II ($\mathbf{B} \approx 1.6\ \hat{\mathbf{z}}\ T$) to III ($\mathbf{B} \approx 1.4\ \hat{\mathbf{z}}\ T$), AFM domains start to form in the $CrBr_3$ layer at the local AFM regions of the moiré supercell (Supplementary Figure S5c-III and Supplementary Video 1), while the $CrI_3$ layer remains FM. The $CrBr_3$ domains reduce the average magnetization below the saturation value and produce a kink in the hysteresis loop.

The AFM domains continue to grow from III ($\mathbf{B} \approx 1.4\ \hat{\mathbf{z}}\ T$) to IV ($\mathbf{B} = 0\ \hat{\mathbf{z}}\ T$), eventually forming three well-established $CrBr_3$ Néel-type skyrmions (Supplementary Figure S5c-IV and Supplementary Video 1) in the local AFM regions at $B = 0\ T$ (IV). The magnetization retains a high remanence at $B = 0\ T$ ($\sim 90\ \%$) since the $CrI_3$ substrate remains FM, while the $CrBr_3$ skyrmions occupy specific regions of the moiré supercell. The reconstruction of the skyrmions at $B = 0\ T$ during the hysteresis manipulation confirms the robustness of the skyrmion lattice and its stability without a permanent magnetic field.

Along the IV-V portion of the hysteresis loop, the external magnetic field points in the negative z-direction, inflating the magnetic skyrmions and reducing the average magnetization. From V to VI, the magnetic skyrmions overgrow rapidly and lose their topological charge, inducing a rapid magnetization reversal in the entire hetero-bilayer at a large coercive field $B \approx -1.865\ T$ (Supplementary Video 1). When the magnetization reversal process is completed, the average magnetization saturates rapidly along $-\hat{\mathbf{z}}$.

A similar hysteresis behavior is observed when the field is swept from $-3\ \hat{\mathbf{z}}\ T$ to $3\ \hat{\mathbf{z}}\ T$. In particular, three $CrBr_3$ Néel-type skyrmions (Supplementary Figure S5c-VII) emerge in the local AFM regions at event VII ($B = 0\ T$). The skyrmion spin textures are opposite at VII and IV (Supplementary Figure S5c-IV and VII).

Importantly, the $CrI_3$ substrate remains FM throughout the hysteresis process except in the close vicinity of the coercive fields (Supplementary Video 1), revealing the solid confinement of the skyrmions to the $CrBr_3$ layer.

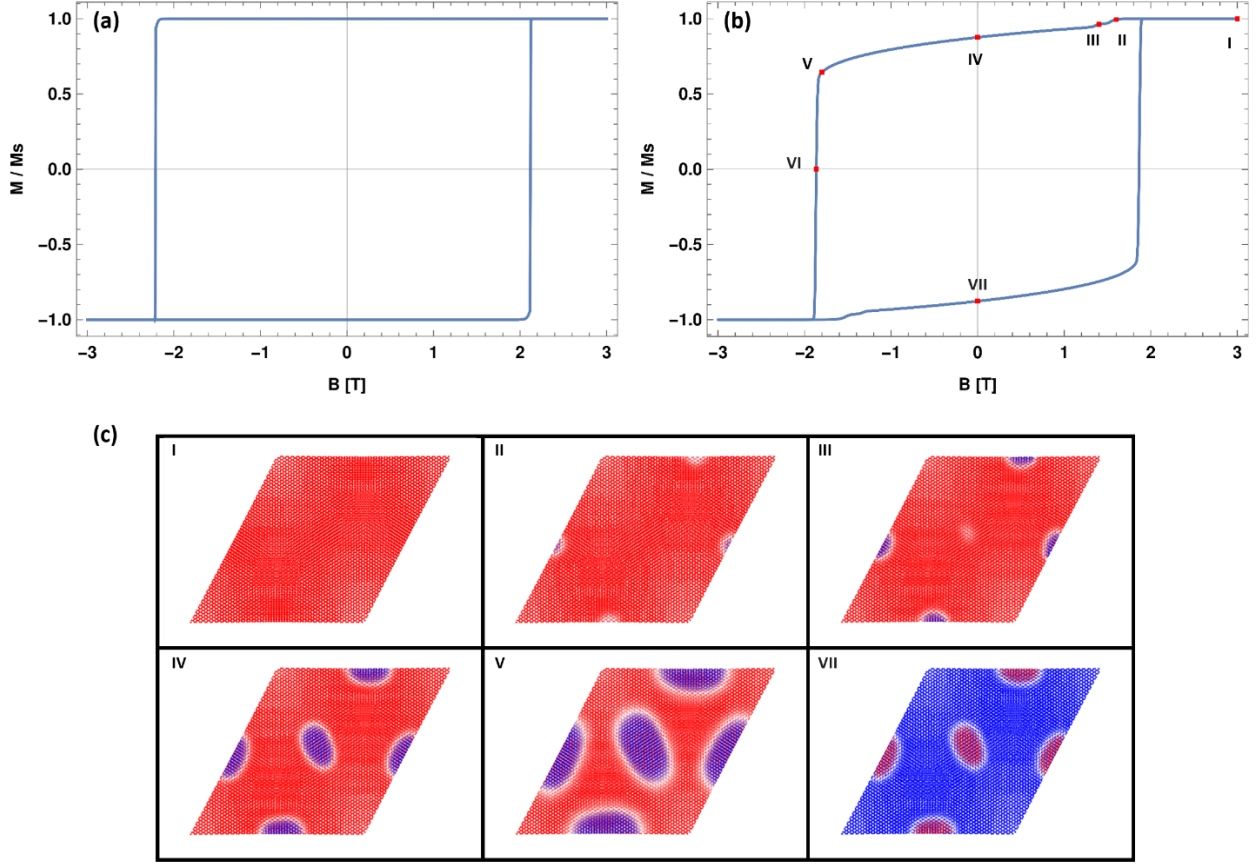

**Figure S5.** Hysteresis loop of a $CrBr_3/CrI_3$ hetero-bilayer in the ferromagnetic (FM) phase (a) and the 3 skyrmions phase (b). The loops were generated using an external magnetic field varied between $\pm 3\,\hat{z}\,T$ in steps of $1\,mT$. Both figures have a Dzyaloshinskii-Moriya interaction (DMI) of $0.3\,meV$, while the twist angle is $6°$ and $1.02°$ in (a) and (b), respectively. The FM hetero-bilayer displays a square-shaped loop (a), indicating a lack of stable magnetic domain formation. In contrast, the 3 skyrmions phase hetero-bilayer displays a fundamentally different behavior (b), with the skyrmion nucleation mediating the hysteresis manipulation (c). The sub-figures in (c) show the $CrBr_3$ spin textures at specific points of the hysteresis loop depicted in figure (b) (see also Supplementary Video 1).